# Transformer ratio enhancement at wakefield excitation in blowout regime in plasma by electron bunch with semi-gaussian charge distribution

*Denys Bondar (1, 2), Vasyl Maslov (1, 2), Irina Levchuk (1), Ivan Onishchenko (1)*
*((1) National Science Center "Kharkov Institute of Physics and Technology", Kharkov, Ukraine*
*(2) V.N. Karazin Kharkov National University, Kharkov, Ukraine)*

vmaslov@kipt.kharkov.ua    bondar@kipt.kharkov.ua

Using 2d3v code LCODE, the numerical simulation of nonlinear wakefield excitation in plasma by shaped relativistic electron bunch with charge distribution, which increases according to Gaussian charge distribution up to the maximum value, and then decreases sharply to zero, has been performed. Transformer ratio, as the ratio of the maximum accelerating field to the maximum decelerating field inside the bunch, and accelerating the wakefield have been investigated taking into account nonlinearity of the wakefield. The dependence of the transformer ratio and the maximum accelerating field on the length of the bunch was investigated with a constant charge of the bunch. It was taken into account that the length of the nonlinear wakefield increases with increasing length of the bunch. It is shown that the transformer ratio reaches its maximum value for a certain length of the bunch. The maximum value of the transformer ratio reaches six as due to the profiling of the bunch, and due to the non-linearity of the wakefield.

The accelerating gradients in conventional linear accelerators are limited to 100 MV/m [1], partly due to breakdown. Plasma-based accelerators have the ability to sustain accelerating gradients which are several orders of magnitude greater than that obtained in conventional accelerators [1, 2]. As plasma in experiment is inhomogeneous and nonstationary and properties of wakefield changes at increase of its amplitude it is difficult to excite wakefield resonantly by a long sequence of electron bunches (see [3, 4]), to focus sequence (see [5-10]), to prepare sequence from long beam (see [11-13]) and to provide large transformer ratio (see [14-20]). Providing a large transformer ratio is also being studied in dielectric accelerators (see [21-26]). In [4] the mechanism has been found and in [27-31] investigated of resonant plasma wakefield excitation by a nonresonant sequence of short electron bunches. Due to the rapid development of laser technology and physics [1, 2, 32–39] laser-plasma-based accelerators are of great interest now. Over the past decade, successful experiments on laser wakefield acceleration of charged particles in the plasma in blowout regime have confirmed the relevance of this acceleration [30–33, 40]. Evidently, the large accelerating gradients in the plasma accelerators in blowout regime allow to reduce the size and to cut the cost of accelerators. Another important advantage of the plasma accelerators in blowout regime is that they can produce short electron bunches with high energy [32]. The formation of electron bunches with small energy spread was demonstrated at intense laser-plasma interactions [41]. Electron self-injection in blowout regime has been studied by numerical simulations (see [37]). Processes of a self-injection of electrons and their acceleration have been experimentally studied in a plasma accelerator [42].

The problem at laser wakefield acceleration is that laser pulse quickly destroyed because of its expansion. One way to solve this problem is the use of a capillary as a waveguide for laser pulse. The second way to solve this problem is to transfer its energy to the electron bunches which as drivers accelerate witness. A transition from a laser wakefield accelerator to plasma wakefield accelerator can occur in some cases at laser-plasma interaction [43].

With newly available compact laser technology [44] one can produce 100 PW-class laser pulses with a single-cycle duration on the femtosecond timescale. With a fs intense laser one can produce a coherent X-ray pulse. Prof. T. Tajima suggested [45] utilizing these coherent X-rays to drive the acceleration of particles. Such X-rays are focusable far beyond the diffraction limit of the original laser wavelength and when injected into a crystal it interacts with a metallic-density electron plasma ideally suited for laser wakefield acceleration [45]. In [46-50] it has shown that at certain conditions the laser wakefield acceleration is added in blowout regime by a beam-plasma wakefield acceleration.

In [51] point self-injected and accelerated electron bunch was observed in blowout regime.

The wakefield excitation in a plasma and its application for particle acceleration avoids the problem of breakdown in the metal structures of accelerators when fields exceeded the value 100 MV/m and creates accelerating gradients which are of considerably higher intensity [2, 4, 52-54].

The efficiency of electron acceleration by a wakefield excited in a plasma by a sequence of electron bunches is determined by the transformer ratio (TR) [16-18, 21-23, 55-70]. The transformer ratio is the ratio of energy acquired by the witness to energy lost by the driver. Approximately, the transformer ratio can be defined as TR=$E_{ac}/E_{dec}$. Where $E_{ac}$ is the maximum accelerating field after the driver bunch (at the end of the first or second bubble). And $E_{dec}$ is the maximum decelerating field inside driver bunch.

Earlier in [60] it was shown that in the linear case, using an Gaussian bunch, the transformer ratio does not exceed TR≤2.

In this work, using a non-linear version of the 2d3v code lcode, numerical simulation of excitation of a non-linear (blowout or bubble mode) wakefield in a plasma by a shaped relativistic electron bunch was performed. Also, the transformer ratio was investigated. In a shaped electron bunch, the charge density along it in the longitudinal direction increases approximately in Gaussian (by cosine) from a zero to maximum, and then abruptly breaks off. The dependence of the accelerating field and TR on the length of the bunch $\xi_b$ is studied when the bunch length $\xi_b$ changes from 0 to the length of the nonlinear wake (bubble), $0<\xi_b<\lambda_{NL}\approx 2\lambda$. Here $\lambda$ is the linear wavelength. It is taken into account that the length of the nonlinear wakefield increases when the length of the bunch increases. In a strongly nonlinear regime, this problem cannot be solved analytically. Therefore, it was investigated using a nonlinear version of the code lcode with a constant charge of the bunch.

For numerical simulation parameters have been selected: relativistic factor of bunch equals $\gamma_b=1000$. The electron plasma frequency is $\omega_{pe}=(4\pi n_r e^2/m_e)^{1/2}$. We consider the bunch, electrons in which are distributed according to Gaussian in the transverse direction along the radius. $\xi=V_b t-z$, $V_b$ is the bunch velocity. Time is normalized on $\omega_{pe}^{-1}$, distance - on $c/\omega_{pe}$, density - on $n_r$, current $I_b$ - on $I_{cr}=\pi mc^3/4e$, fields – on $(4\pi n_r c^2 m_e)^{1/2}$.

We use the cylindrical coordinate system (r, z) and draw the plasma and beam densities and longitudinal electric field at some z as a function of the dimensionless time $\tau=\omega_p t$.

The longitudinal coordinate $\xi=z-V_b t$ is normalized on $\lambda/2\pi$. The values of the $E_z$, $F_r$ and $H_\theta$ are normalized on $mc\omega_{pe}/e$. Where e, m are the charge and mass of the electron, c is the light velocity, $\omega_{pe}$ is the electron plasma frequency.

We do not take into account the longitudinal dynamics of the bunches, because at the times and energies of the beam according to

$$\frac{dV_z(r)}{dr} \propto \frac{1}{\gamma_b^3}, \quad \frac{dV_r(r)}{dr} \propto \frac{1}{\gamma_b}$$

radial relative shifts of beam particles predominate. $V_z$, $V_r$ are the longitudinal and radial velocities of the electron bunches, $\gamma_b$ is the relativistic factor of the bunch.

The aim of the paper is the demonstration by the numerical simulation that the transformer ratio – an important value in the wakefield method of acceleration of electron bunches, can be increased by a factor of three due to the profiling of the electron-driver-bunch and due to the non-linearity of the excited wakefield.

## INVESTIGATION OF THE TRANSFORMER RATIO

We consider wakefield excitation in the plasma by the bunch near the injection boundary, since the bunch is deformed when the penetration into the plasma is deep. The main purpose of this work - to consider factors which can increase the transformer ratio when the bunch excites wakefield in plasma. It is also an important task to search for the optimal length of the bunch to obtain the highest transformer ratio.

In Fig. 1 the dependence of the value of the transformer ratio on the bunch length is shown for the case of the first bubble.

The value of the transformer ratio increases almost linearly with increase of the length of the bunch, until the length of the bunch reaches $1.125\lambda$. But then after a local maximum at a bunch length of $1.125\lambda$ the transformer ratio increases with bunch length increase. The largest value of the transformer ratio is achieved when the bunch length equals $7\lambda/4$. I.e. the largest transformer ratio is achieved through the interval of the length of the bunch, approximately equal $\lambda/2$. One can see in Fig. 1 that the transformer ratio reaches maximum value $TR^{1st}\approx 5.25$ when the bunch length equals to $7\lambda/4$ for the first bubble in the nonlinear regime for profiled bunch. Further increase of a bunch length leads to a decrease of the transformer ratio.

Thus, we can state that the length of the bunch $7\lambda/4$ is the optimal length from the point of view of the efficiency of electron acceleration by the excited wakefield, namely by the wakefield at the end of the first wake bubble. $\lambda$ is the wavelength of the linear wakefield.

Further, we consider the transformer ratio for the second bubble. One can see in Fig. 1 that in this case the transformer ratio behaves in a similar way, as in the case of the first bubble. First, TR increases almost linearly with increasing of the length of the bunch, until the bunch length reaches the value of $3\lambda/4$. After that one can see the maximum when the length of the bunch reaches $7\lambda/4$ (the interval from the first maximum is $\lambda$). When the length of the bunch reaches $1.125\lambda$, in the both cases of the first and the second bubble, small (relative to the main maximum) jump of the value of the transformer ratio is observed.

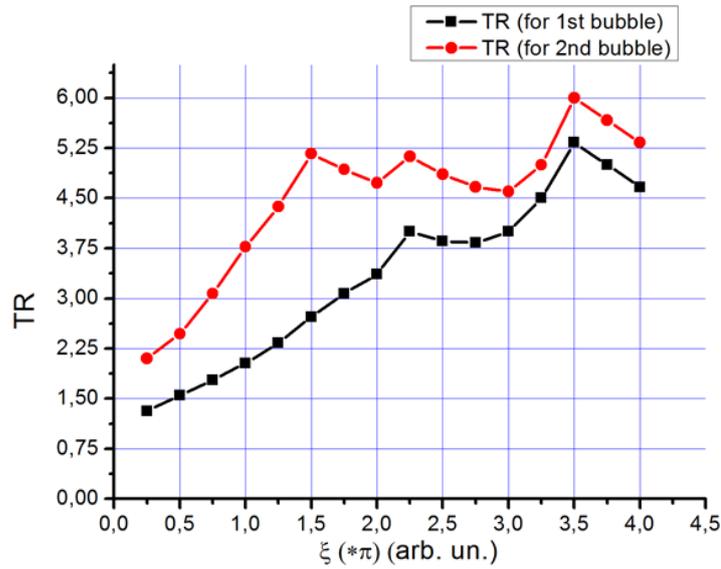

Fig. 1. Dependence of the transformer ratio on the length of the bunch for the 1st and 2nd bubble (normalized by the wavelength of a linear wakefield)

Further, after the TR maximal value at length of the bunch $7\lambda/4$ the transformer ratio decreases. For the second bubble, the maximum transformer ratio for the bunch length $7\lambda/4$ is equal to $TR^{2nd} \approx 6.00$. For the subsequent (after the first) bunches, the transformer ratio can increase at certain conditions. This can occur due to the accumulation (summation) of the wakefield at approximately the same the decelerating field for all bunches.

Moreover, it is remarkable that the maximums of the transformer ratio are observed at the same length of the bunch after the first and the second bubbles. This leads to the possibility to accelerate two bunches: one bunch at the end of the first bubble, and the second bunch at the end of the second bubble, placing them to the maximum accelerating fields at the bunch length equal to $7\lambda/4$.

## INVESTIGATION OF THE ACCELERATING AND DECELERATING FIELDS

Further, the dependence of the accelerating field from the length of the bunch in the nonlinear regime was studied for the shaped driver-bunch (Fig. 2). It is observed that the amplitude of the excited nonlinear wakefield decreases (in absolute value) when the length of the bunch increases, similarly to the case of a linear wakefield with an unformed bunch investigated by other authors [55].

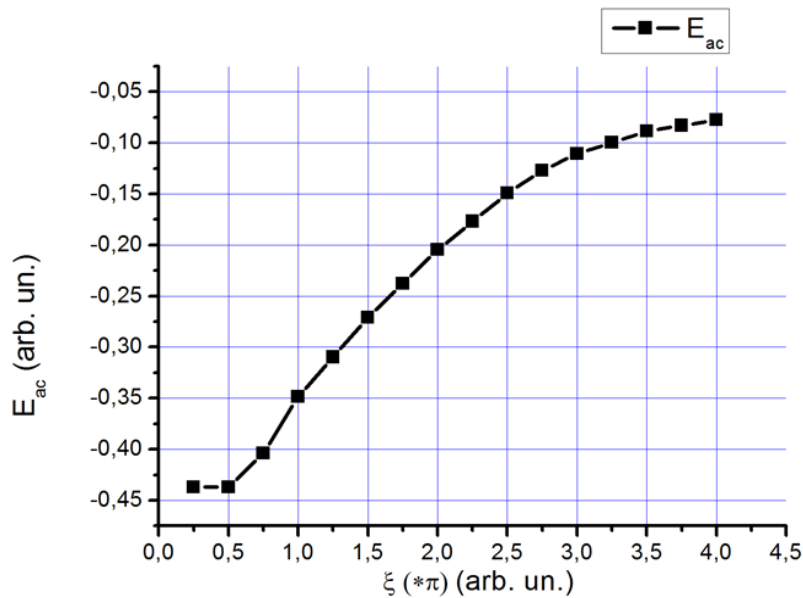

Fig. 2. Dependence of the accelerating field on the length of the bunch for the 1st bubble.

This is determined by the finite length of the considered bunch. Indeed, each point of the bunch excites a field whose distribution can be approximately described by a semi-cosine. However, since the fields are excited at different points, some of them are in antiphase and suppress each other. Therefore, the amplitude of the accelerating field decreases with increasing length of the bunch.

In the case of the second bubble (Fig. 3), a similar dependence is observed: in absolute value, the amplitude of the accelerating field decreases. However, in the case of the second bubble, the amplitude of the accelerating field is initially larger than the amplitude of the accelerating field in the case of the first bubble. Strictly speaking, this leads to an increase of the maximum value of the transformer ratio in the case of the second bubble. The excess of the maximum accelerating field after the second bubble over the accelerating field after the first bubble can be explained by the inertness of the plasma electrons, which received a pulse from the driver bunch; and by influence of the space charge of the driver bunch.

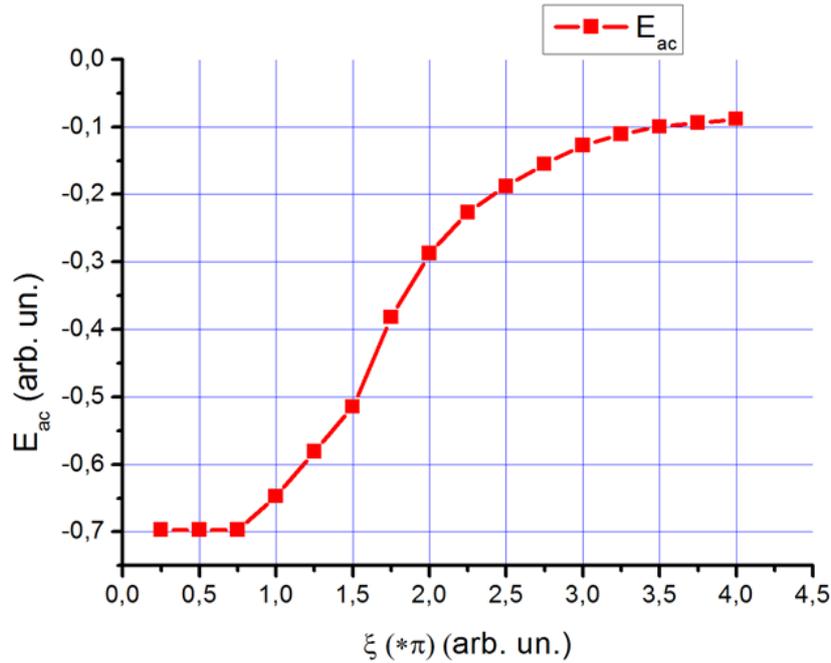

Fig. 3. Dependence of the accelerating field on the length of the bunch for the 2nd bubble.

In our case, when the length of the bunch increases with a fixed charge of the bunch, i.e. with a fixed number of electrons in the bunch, when electron density in the bunch $n_b$ decreases, the bubble lengthens Fig. 4. **(a) (b)**

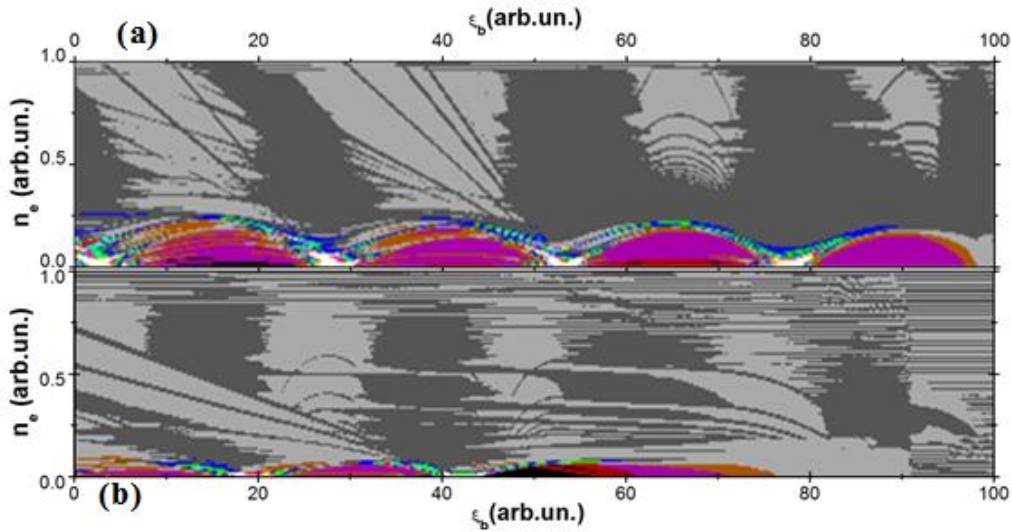

Fig. 4. Spatial distribution of the density of plasma electrons.
The figure above (a) is for a bunch length equal to $0.25*\pi$. The figure below (b) is for the length of the bunch, equal to $3.75*\pi$.

In addition, the dependence of the decelerating field on the length of the bunch was investigated (Fig. 5).

One can see from Fig. 5 that the dependence of the decelerating field on the length of the bunch is a function that decreases monotonically with increasing of the length of the bunch.

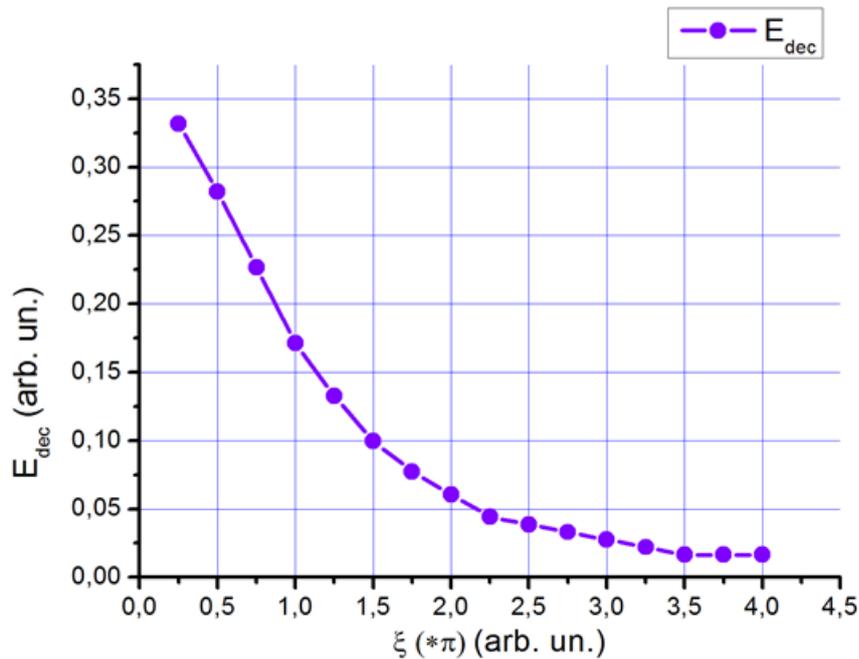

Fig. 5 Dependence of the decelerating field on the length of the bunch

## CONCLUSION

Thus, we can state the following. In this paper, it was demonstrated by numerical simulation that the transformer ratio at the wakefield excitation by a bunch of relativistic electrons increases due to the profiling of the bunch, and also due to the nonlinearity of the excited wakefield. The value of the transformer ratio after the second bubble exceeds the transformer ratio after the first bubble. It is shown that for certain values of the length of the bunch, the transformer ratio reaches a maximum value exceeding the transformer ratio in the linear case in the absence of shaping of the bunch. The dependence of the accelerating and decelerating fields on the length of the bunch were also investigated and they were established that the obtained dependences agree with the theoretical assumptions.